\documentclass[10pt, conference]{IEEEtran}
\usepackage[latin1]{inputenc}
\usepackage{adjustbox}
\usepackage[normalem]{ulem}
\useunder{\uline}{\ul}{}
\usepackage{cite}

\ifCLASSINFOpdf
\else
\fi
\usepackage[hyphens]{url}

\usepackage{graphicx}
\usepackage{multirow}
\usepackage{stfloats}

\usepackage{pdflscape}
\usepackage{array,booktabs}
\usepackage[none]{hyphenat}
\begin{document}

\newcommand{\ra}[1]{\renewcommand{\arraystretch}{#1}}
\newcolumntype{L}{>{\raggedright\arraybackslash}p}
\newcolumntype{C}{>{\centering\arraybackslash}p}
\newcolumntype{F}{>{\arraybackslash}p}
\newcolumntype{R}{>{\raggedleft\arraybackslash}p}

\title{Reflections on Cyberethics Education for Millennial Software Engineers}

\author{\IEEEauthorblockN{Claudia de O. Melo}
\IEEEauthorblockA{Faculty of Technology\\
University of Bras\'{\i}lia\\
Bras\'{\i}lia, Brazil\\
Email: claudiam@unb.br}
\and
\IEEEauthorblockN{Thiago C. de Sousa}
\IEEEauthorblockA{Urbanism and Technology Center\\
State University of Piau\'{\i}\\
Teresina, Brazil\\
Email: thiago@uespi.br}
}

\maketitle

\begin{abstract}
Software is a key component of solutions for 21st Century problems. These problems are often ``wicked", complex, and unpredictable. To provide the best possible solution, millennial software engineers must be prepared to make ethical decisions, thinking critically, and acting systematically. This reality demands continuous changes in educational systems and curricula delivery, as misjudgment might have serious social impact. This study aims to investigate and reflect on Software Engineering (SE) Programs, proposing a conceptual framework for analyzing cyberethics education and a set of suggestions on how to integrate it into the SE undergraduate curriculum.
\end{abstract}

%software engineering (326), software engineer (163), millennial software engineer (126), ethical issue (120), software development (90), computer ethic (80), artificial intelligence (80), distributed system (80), non technical component (79), human computer interaction (79), work load (70), cloud computing (70), undergraduate degree program (63), digital business model (63), social scientist (60), disclosive method (60), big data (60), total content (60), curriculum guideline (60), conceptual framework (60), database system (60), data science (50), computer technology (50), software architecture (50), secure software engineering (47), ufc software requirement (47), brazilian software engineering (47), software requirement (46), software design (45), moral issue (40)

% \fi
\IEEEpeerreviewmaketitle
\section{Introduction}
% Problem domain (2 paragraph)
% General gap between digital age wicked problems and how millennials are prepared in the SW Engineering undergraduate courses (new education is general needed and how this affects the SE curriculum).
% * Specific Problem: Solving wicked problems require ethics understanding. How ethics for 21st century would change the current SE curriculum
% * How we are going to address the problem: 
%  * analyze the most advanced reference frameworks on 21st century competences framework (general view to deal with wicked problems) + digital information ethics (what's needed to be ethical?) +  VERSUS SE Engineering curriculum to find gaps)
%  * propose how to fill the gaps using a concrete example, the BR SE Engineering curriculum
% the ACM/IEEE Software Engineering Code of Ethics and Professional Practice/Ethics of Information and Communication Technologies.
% analyze top five BR SE undergrad courses looking for evidences of digital age educational components (qualitative methods --)   

The current technological revolution, being a critical engine of the digital transformation of the economy, is impacting all disciplines and industries. The so-called ``Digital Economy" is considered the single most important driver of innovation, competitiveness and growth of countries \cite{EC2016}. In this context, it is also a consensus among nations that taking advantage of technology to advance social  and  economic inclusion, and to promote sustainability and peace, is paramount and will demand a ``transformation of societies"\cite{UN2016}.

%Because of the intrinsic role of technology in the Digital Economy, there is a need for better understanding technology nature and impacts. Technological change is not neutral and is, in essence, disruptive. It creates, in the short term, winners and losers. While disruptive technologies will be critical to a transformation towards sustainable development, their benefits may disproportionately go to people in the countries that innovate or to a small fraction of the population \cite{UN2016}. 

Software engineers investigate problems, and propose and develop software to tackle such societal challenges. They are creating the foundations that enable and govern our online and increasingly our offline lives, from software-controlled cars to digital content consumption. In fact, software helps shape, not just reflect, our societal values \cite{Narayanan2014}.

Software is a key component of solutions for 21st Century problems  \cite{Easterbrook2014}. These problems are complex and unpredictable, which is typical of ``wicked" problems \cite{Rittel1973}. They are difficult to define and are never entirely solved, as improvements can always be made \cite{Dyba2014}. 

To be able to work on problems not seen before, while applying the right knowledge and judgment, software engineers must think and act  more systemically and more adaptively, so they can build up experience over time to make sense of world challenges.

As the complexity of problems continues to increase, the complexity of software  intensive  systems  or  systems
of  systems also increases \cite{Boehm2006}. Therefore, there  are  serious challenges to  educating millennial software engineers, which requires continuous innovation in educational systems and curricula delivery\cite{Boehm2006,UN2016}.

%The failure to think systemically is a critical weakness in our understanding of wicked problems. This trap arises because of how we educate ICT professionals \cite{Easterbrook2014}. 
Moreover, the design of software systems comes with a special set  of responsibilities to society that are much broader than those described in existing codes of ethics for computing professionals \cite{Becker2015}. While the potential return on investment in technology is usually high, the increasing pace of technological innovation raises ethical questions about its development and use \cite{UN2013}. Information and Communications Technologies (ICT) bring `predictable' and `less predictable' ethical issues \cite{Stahl2013}. Unintended consequences of technology \cite{Tenner1997} need to be investigated and a precautionary principle \cite{Riordan2013} applied.  

%confirmar se millenials são considerados the current workers generation
Do millennial software engineers, the current dominant generation of workers, understand the ethical choices and related unintended consequences that the solutions for the 21st century might generate? Are they prepared to investigate and co-design solutions with other stakeholders to ensure better solutions for all? From algorithms, data science, AI, cloud computing to digital business models and services design, there are a number of ethical decisions a software engineer is (or not) going to make while designing systems.

This paper aims at investigating these questions about how millennial software engineers are trained to deal with computer ethics issues. We developed a conceptual framework based on Brey's ``disclosive" method for cyberethics, as it provides the major components for cyberethics decision-making, and the most important 21st Century cybertechnologies to support our investigation. We analyze the ACM/IEEE Software Engineering Guidelines and the curricula of the top two Brazilian SE undergraduate programs. We end the paper with a set of suggestions that might be integrated into the SE undergraduate curriculum, as well as conclusions and recommendations for future studies.

\section{Cyberethics}
\subsection{Global and Regional contexts}
Edward Snowden's revelations on the Five-Eyes mass surveillance alliance, starting in 2013, are examples of how technology can be used to concentrate economic power and create global monopolies \cite{Melo2016}. 
%The espionage depends on the software code (algorithms) and the witting or unwitting cooperation of Telecom corporations. Individual users may play a part, but their role is mostly not conscious \cite{Lyon2013}.

%As Greenwald et al. \cite{Greenwald2013} describes, espionage targets the communications of everyone, then filters, analyzes, measures them and stores them for periods of time simply because it is the easiest, most efficient and most valuable way of achieving these ends.

These revelations had profound impacts on society's perception of ethical issues in the digital age, increasing the importance of cyberethics. Despite that technology has great potential of freeing people from manual and repetitive labor, it also brings other troubling concerns, requiring from human understanding and accountability \cite{Knightfoundation2017}. 

%As mentioned by Jonathan Zittrain, co-founder of the Berkman Klein Center and Professor of Law and Computer Science at Harvard University, ``a lot of our work in this area will be to identify and cultivate technologies and practices that promote human autonomy and dignity rather than diminish it" \cite{Knightfoundation2017}.

Considering this context, countries are launching large-scale initiatives focused on addressing cyberethic issues. For instance, in the United States, several foundations and funders recently announced the Ethics and Governance of Artificial Intelligence Fund, which will support interdisciplinary research to ensure that AI develops in a way that is ethical, accountable, and advances the public interest \cite{Knightfoundation2017}. 

%The Harvard Berkman Klein Center and the MIT Media Lab are the ``anchor academic institutions" for the initiative, bridging the gap between disciplines and connecting human values with technical capabilities through activities, research, tools, and prototypes. 
%They state that ``it is imperative that AI research and development be shaped by a broad range of voices - not only by engineers and corporations - but also social scientists, ethicists, philosophers, faith leaders, economists, lawyers, and policymakers".
%They have noted that the potential for empowerment through the use of robotics is nuanced by a set of tensions or risks relating to human safety, freedom, health, privacy, integrity, dignity, autonomy, personal data protection, data ownership, the freedom to obtain or transfer information, freedom of expression and freedom to conduct a business
The European Union is currently discussing the creation of a European agency for robotics and AI. They aim to preserve human dignity and integrity while developing artificial intelligence and robots. \cite{EUParliament2016}. 

%They are now developing policies and a legal framework that defines what `smart autonomous robots' are, and also an advisory code of conduct for robotic engineers aimed at guiding the ethical design, production and use of robots.

In Brazil, Marco Civil (the Brazilian civil rights framework for the Internet) aims to protect human rights, including ensuring freedom of speech and expression, protecting privacy and personal data, ensuring equitable access to information, and promoting an open, competitive online marketplace, partly by guaranteeing net neutrality \cite{Melo2016}. 
%While some of Marco Civil's regulations are still being discussed, companies are already infringing privacy and net neutrality existent regulations on pursuing activities to increase economic profits.

\subsection{What is cyberethics?}
% Is Digital Economy sustainable? (an ethical perspective)
% The role of Software Engineers ethics in the Digital Economy
% How are we educating SW Engineers (Millennials) for the future
According to Moor \cite{Moor1985}, computer ethics is the ``analysis of the nature and social impact of computer technology and the corresponding formulation and justification of policies for the ethical use of such technology".

%Many authors have used the term ``computer ethics" to describe the field that examines moral issues pertaining to computing and information technologies; while others use ``information ethics" to frame ethical concerns around the flow of information that is enhanced or restricted by computer technology \cite{Tavani2016}. There are also emergent fields of ``agent ethics", ``bot ethics", ``robo-ethics", and ``machine ethics" that overlap with concepts of cyberethics \cite{Tavani2016}.

Tavani \cite{Tavani2016} argues that \textbf{cyberethics} is a more appropriate and accurate term, connoting the social impact of computers and cybertechnology in a broad sense and not merely the impact of that technology for computer professionals. He defines cyberethics as the ``study of moral, legal, and social issues involving cybertechnology.'' Cybertechnology, in turn, comprises the entire range of computing and communication systems, from stand-alone computers to privately owned networks and to the Internet itself.

%confirmar em tavani
%Thus, cyberethics in the broadest sense can be understood as the branch of applied ethics which studies and analyzes social and ethical impacts of information technology.

%Although it may be difficult to prove conclusively whether or not cybertechnology has generated any new or unique ethical issues \cite{}, we must not rule out the possibility that many of the controversies associated with this technology warrant special consideration from an ethical perspective \cite{}. 

Moor \cite{Moor1985} explains that computers are essentially a malleable, universally applicable tool, so the potential applications for human action and consequent ethical issues are novel and almost limitless \cite{Mingers2010,Tavani2016}. Some of these actions might generate what Moor calls ``\textit{policy vacuums}", because we have no explicit policies or laws to guide new choices that are only possible through cybertechnology. These vacuums, in turn, need to be filled with either new or revised policies \cite{Tavani2016}.

According to Patrignani \cite{Patrignani2014}, we are living in a policy vacuum era and nobody questions technologies. Scenarios for their use change and evolve rapidly and there are no policies addressing these new situations. Therefore, we might assume that whoever is designing new technologies is going to be especially responsible for any unintended consequence.

%Viewing computer ethics issues in terms of policies is useful, because policies have the right level of generality to consider when we evaluate the morality of conduct. 
Policies are ``rules of conduct, ranging from formal laws to informal, implicit guidelines for actions" \cite{Moor1985}. Policies \textit{can range from formal laws to informal guidelines} \cite{Tavani2016}.
%Moor also explains that policies can have ``justified exemptions" because they are not absolute. However, they usually imply a certain ``level of obligation" within their contexts
%What action is required to resolve a policy vacuum when it is discovered? Initially, a solution to this problem might seem quite simple and straightforward. We might assume that all we need to do is identify the vacuums that have been generated and then fill them with policies and laws.
However, this will not always work, because sometimes the new possibilities for human action generated by cybertechnology also introduce ``\textit{conceptual muddles}" \cite{Moor1985}. In these cases, we must first eliminate the muddles by clearing up certain conceptual confusions before we can frame coherent policies and laws \cite{Tavani2016}.

In addition, Brey \cite{Brey2000, Brey2010} believes that because of embedded biases in cybertechnology, the standard applied-ethics methodology is not adequate for identifying and acting on cyberethics issues. Brey argues that the standard ethics method tends to focus almost solely on the \textit{uses} of technology \cite{Tavani2016}. The standard method fails to pay sufficient attention to certain features and practices involving the use of cybertechnology that have moral import but that are not yet known \cite{Brey2000}. 
%Brey refers to such practices as having ``morally opaque" (or morally nontransparent) features, which he contrasts with ``morally transparent" features \cite{Brey2000}.

Many practices involving computer technology are morally nontransparent because they include operations of technological systems that are very complex and difficult to understand for laypersons and that are often hidden from view for the average user \cite{Brey2010}. They also involve distant actions over computer networks by system operators, providers, website owners and hackers and remain hidden from view from users and from the public at large \cite{Brey2010}. 

\subsection{Making cyberethics decisions: Brey's ``disclosive" method}
%Brey's ``disclosive" method for cyberethics
To address aforementioned caveats on cyberethics decision-making, Brey proposes a ``disclosive" method for cyberethics \cite{Brey2000}. The aim of disclosive ethics is to identify such morally opaque practices, describe and analyze them, so as to bring them into view, and to identify and reflect on any problematic moral features of cybertechnologies \cite{Brey2010}. 

Brey describes the methodology for computer ethics: it must first identify, or ``disclose'', features that, without proper probing and analysis, would go unnoticed as having moral implications. Therefore, we need computer scientists (or software engineers) because they better understand computer technology (as opposed to philosophers and social scientists) \cite{Brey2000}. 

Social scientists are also needed to evaluate system designs and make them more user-friendly. Philosophers can determine whether existing ethical theories are adequate to test the newly disclosed moral issues or more theory is needed. Finally, computer scientists, philosophers, and social scientists must cooperate in applying ethical theory in deliberations about moral issues \cite{Brey2000}.
%Brey's disclosive method is multidisciplinary because it requires the collaboration of: computer scientists, philosophers, social scientists. It is also multi-level because the method for conducting computer ethics research requires three levels of analysis: disclosure level, theoretical level, application level.

Tavani \cite{Tavani2016} summarizes Brey's disclosive method in the three following steps. 
\begin{description}
\item[Step 1] Identify a practice involving cyber-technology, or a feature in that technology, that is controversial from a moral perspective.1a. Disclose any hidden (or opaque) features or issues that have moral implications. 1.b If the ethical issue is descriptive, assess the sociological implications for relevant social institutions and socio-demographic and populations. 1c. If the ethical issue is also normative, determine whether there are any specific guidelines, that is, professional codes that can help resolve the issue. 1d. If the normative ethical issues remain, proceed to Step 2.
\item[Step 2] Analyze the ethical issue by clarifying concepts and situating it in a context. 2a. If a policy vacuum exists, proceed to Step 2b; otherwise continue to Step 3. 2b. Clear up any conceptual muddles involving the policy vacuum and proceed to Step 3.
\item[Step 3] Deliberate on the ethical issues. The deliberation process requires two stages: 3a. Apply one or more ethical theories to the analysis of the moral issue, and then go to step 3b. 3b.Justify the position you reached by evaluating it against the rules for logic/critical thinking. 
\end{description}

% We need to be thinking about every decision and every action contributing to a system operating under ethical principles. Sustainability provides a framework for expanding ethical reasoning to a complex world. COMPUTING EDUCATION FOR SUSTAINABILITY— WHAT GIVES ME HOPE?

\section{Conceptual framework for analyzing millennial software engineers education for cyberethics}
\begin{figure*}[bth]
      \begin{center}
      \includegraphics[scale=0.67]{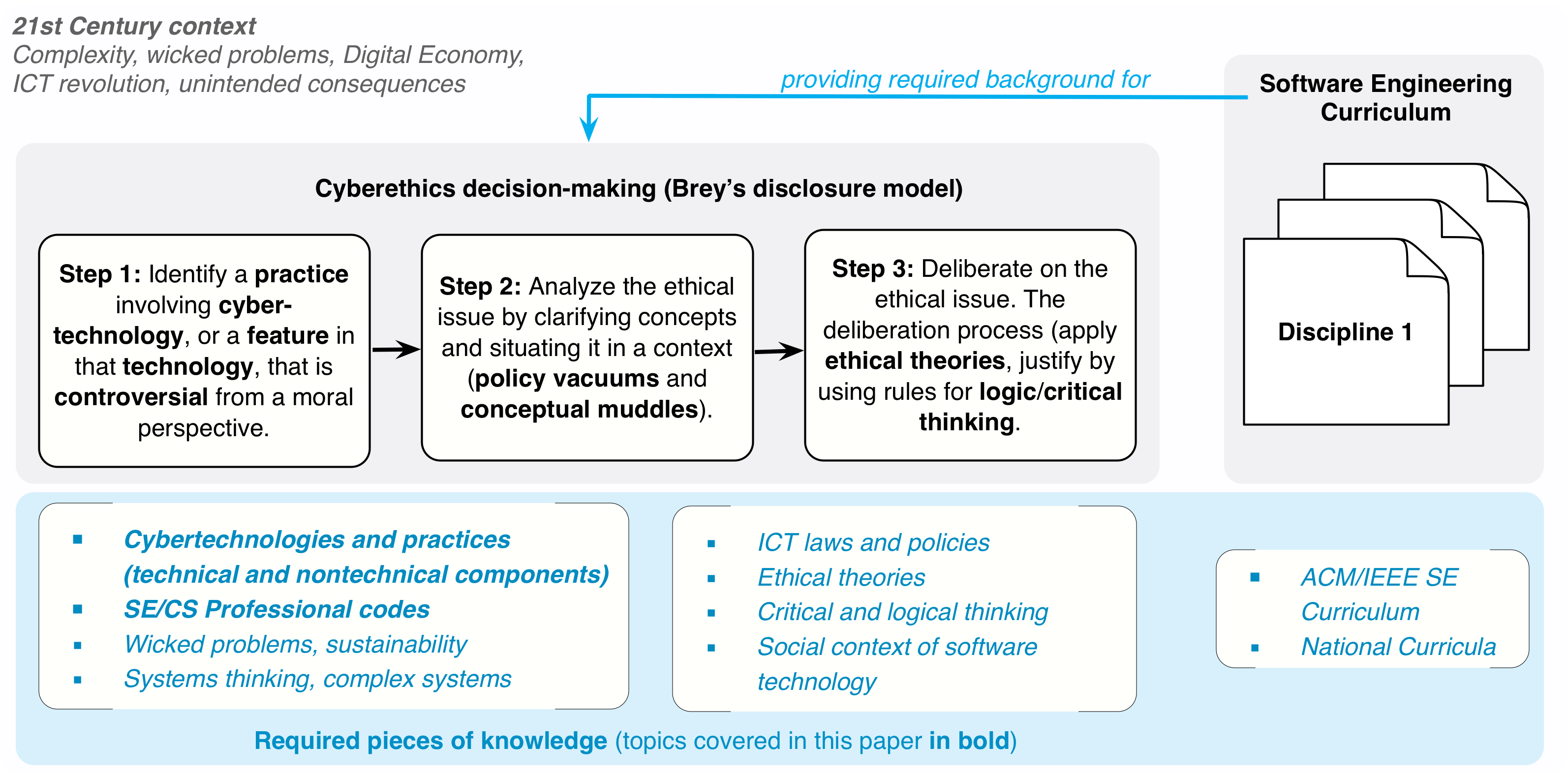}
      \end{center}
          \vspace{-1.5em}  
%      \caption{(a) Thematic network analysis stages, adapted from \citep{Attride-Stirling01}. (b) Example of Stage ``Reduction or Breakdown of Text'}
\caption{Conceptual framework for analyzing cyberethics of Software Engineering curriculum }
      \label{fig:conceptualframework}
 \end{figure*}

To be able to analyze millennial software engineers' education under the lens of cyberethics, we derived a conceptual framework  from  1) Brey's  ``disclosive" method on  cyberethics decision-making and  2)  the  required pieces of knowledge to apply it, which should be provided by 3) a certain Software Engineering curriculum. Figure \ref{fig:conceptualframework} illustrates our conceptual framework.

Despite that we provide a number of required knowledge pieces, this paper focuses only on cybertechnologies and practices, as well as the usage of already available SE/CS professional codes.  We   reviewed the   literature  on   cybertechnologies to  identify  the  most  relevant  technical  and  nontechnical components that are crucial for Software Engineers in the 21st century,  as  they  might  exhibit  controversial issues  from  a moral perspective. We took advantage of our previous 15 years experience in global software development companies to refine and complement this list of relevant cybertechnologies. 

%We illustrate knowledge pieces required to accomplish each of the Brey's model steps. The conceptual framework aims at analyzing any Software Engineering Curriculum. Thus, we give examples of the ACM/IEEE SE Curriculum and National Curricula. 

% adapted from A Multi-level Perspective for the Integration of Ethics, Corporate Social Responsibility and Sustainability (ECSRS) in Management Education
%We adopt a multi-discipline perspective on integrating cyberethics into the SE curriculum because xxx
%21st-century most important cybertechnologies features and practices \cite{ETICA2011}
%To be able to discuss the findings and provide a set of recommendations for integrating cyberethics into the SE curriculum, we considered:
%Multi-discipline perspective on integrating cyberethics into the SE curriculum
% Pathways towards responsible ICT Innovation ETICA2011

In 2011, the Etica project, funded by the European Commission, analyzed 100 technologies, 70 application examples and 40 artifacts to synthesize the emergent ICTs coming in 10 to 15 years. The emergent technologies are high-level socio-technical systems that have potential to affect the way humans interact with the world \cite{ETICA2011}.

They highlighted 11 ICTs: Affective Computing, Ambient Intelligence, Artificial Intelligence, Bioelectronics, Cloud Computing, Future Internet, Human-machine symbiosis, Neuroelectronics, Quantum Computing, Robotics, and Virtual/Augmented Reality.

We found that some additional components must be part of the list of cybertechnologies. For instance, despite that Data Science, Social Media, and Mobile are not emergent, they are structuring technologies for the necessary hyper-connectivity that enables the Digital Economy. Rapid advances in technology can boost whole business models \cite{Ramesh2016}, which means that digital business models can have enormous impact on society and the economy. As customers expect increasingly faster, cheaper, and better products and services, practices involving Design Thinking and Experience, Product and Services Design become crucial for software companies competing against the already known experiences customers have with the Facebooks, Apples, and Amazons of the world \cite{Ramesh2016}. Finally, software security and technological social impact are must-have components. 

After removing ICTs that we considered too futuristic (e.g., neuroelectronics or human-machine symbiosis), we ended with a list of 9 crucial cybertechnologies for the 21st century that are relevant to this study. For didactic purposes, we divided the critical into two groups: technical and non-technical components. The technical components are: 1) Data Science; 2) Cloud Computing; 3) Algorithms; 4) Artificial Intelligence; Robotics; 5) Internet; Internet of Things; Mobile; Social Media; 6) Secure Software Engineering. The non-technical components are: 7) Experience, Product and Services Design; Design Thinking; 8) Digital Business Models; Economics; 9) Software and Society; Cyberlaw.

% with the world. 
% These were synthesised into the following list of emerging ICTs. The term "emerging ICT" is used for
% high-level socio-technical system that has the potential to significantly affect the way humans interact
% with the world. They are deemed to be emerging if they are likely to be socially and economically
% relevant in the coming 10 to 15 years. They highlighted the following 11 ICTs: affective Computing, Ambient Intelligence, Artificial Intelligence, Bioelectronics, Cloud Computing, Future Internet, Human-machine symbiosis, Neuroelectronics, Quantum Computing, Robotics, and Virtual/Augmented Reality.

%COMPUTER ETHICS: A SYLLABUS FOR TEACHING ETHICS IN COMPUTER SCIENCE

Table \ref{tab:tech} presents the nine suggested components, a summary of their moral controversies and key references. 
\begin{table*}[tbh]\centering
\ra{0.7}
\caption{21st century cybertechnologies controversial issues}
\label{tab:tech}
\footnotesize
\begin{tabular}{@{}L{4cm} L{11.5cm} C{1.5cm}@{}}
\toprule
\textbf{Cybertechnologies}& \textbf{Controversial practices or features, from a moral perspective}& \textbf{References}   
\\ \midrule
Data Science & Discoveries in data mining, propensity and group privacy.& \begin{tabular}[c]{@{}l@{}} \cite{zwitter} \\ \cite{Fairfield2014} \end{tabular}                               \\ \midrule
Cloud Computing & Consumer privacy, reliability of services, data ownership and technology neutrality.& \begin{tabular}[c]{@{}l@{}}\cite{Timmermans}\\ \cite{ deBruin2017}\end{tabular} 
\\ \midrule
Algorithms & Inconclusive evidence leading to unjustified actions, inscrutable evidence leading to opacity, misguided evidence leading to bias, unfair outcomes leading to discrimination, transformation effects leading to challenges for autonomy and traceability leading to moral responsibility. & \begin{tabular}[c]{@{}l@{}}\cite{brent} \\ \cite{Kraemer2011} \end{tabular}  
\\ \midrule
Experience, Product and Services Design; Design Thinking & Participatory design conflict, amount of time and energy required by successful integration into design team and tension between firm grounding contexts and abstracted model of design.            & \begin{tabular}[c]{@{}l@{}}\cite{shilton} \\ \cite{salvo}\end{tabular}          
\\ \midrule
Artificial Intelligence; Robotics & Machine learning, bias in natural language processing and robots as sexual partners, caregivers, and servants. & \begin{tabular}[c]{@{}l@{}} \cite{AndersonA07} \\  \cite{VeruggioO08}  \end{tabular}         \\ \midrule
Digital Business Models; Economics & Intellectual property rights, economic market impact and customer relationship. & \begin{tabular}[c]{@{}l@{}} \cite{johnson2016ethics} \\ \cite{trevino2010managing} \end{tabular}                         \\ \midrule
Internet; IoT; Mobile; Social & Individual privacy preferences, access controls, emergent social conventions and infrastructures for government surveillance.                 & \begin{tabular}[c]{@{}l@{}}\cite{saty}\\ \cite{ling}\end{tabular}               
\\ \midrule
Software and Society; Cyberlaw & Cryptocurrency, net neutrality, proprietary code and content and freedom of speech.& \begin{tabular}[c]{@{}l@{}}\cite{Dudley} \\ \cite{Ermann}\end{tabular}          
\\ \midrule
Secure Software Engineering & Purposeful human errors injection, software piracy and software development for espionage, extortion, vandalism and theft. & \begin{tabular}[c]{@{}l@{}}\cite{Whitman} \\ \cite{VonSolms}\end{tabular}       \\ \bottomrule
\normalsize
\end{tabular}
\end{table*}

\section{An Analysis of the SE Curriculum}
In this section, we analyze the ACM/IEEE Curriculum Guidelines for Undergraduate Degree Programs in Software Engineering \cite{curricula}, as well as the top two Brazilian Software Engineering Programs for evidence of ethics as an interdisciplinary approach addressing the needs of the future. More specifically, we searched for topics related to ethics in the ACM/IEEE Software Engineering Education Knowledge (SEEK) and in the two Brazilian SE undergraduate courses to check if the teaching of Ethics is one given appropriate attention in millennial software engineers' education. For this purpose, we performed a search for some Ethics-related topics presented in syllabi of this discipline, such as ``Ethics", ``Society", ``Sustainability", ``Environment", ``General Systems Theory", ``Complex and Adaptive System", ``Law", ``Legal", ``Social" and ``Humanity".       

Ethics is one of the most valued concepts presented in the ACM/IEEE Computing Curricula. In addition to explicitly citing the Software Engineering Code of Ethics and Professional Practice \cite{Gotterbarn}, the Curriculum Guideline number 15 states that ``Ethical, legal, and economic concerns and the notion of what it means to be a professional should be raised frequently." Moreover, the Ethics word appears 19 times. 

The proposed SEEK, which is inspired in the SWEBOOK \cite{IEEEComputerSociety:2014} and describes the body of knowledge that is appropriate for software engineers, suggests 467 hours of fundamental content  for  the  design,  implementation,  and  delivery  of the  educational units  that  make up  a  software engineering curriculum. However, there was only one Ethics-related topic, named ``Professionalism" with 6 hours, which corresponds to 1.3\% of the total content, which seems to be an inconsistency. Although    there  are    no    Curriculum    Guidelines   for Undergraduate  Degree  Programs  in  Software  Engineering provided by the Brazilian Computer Society, the Computing Curriculum Manuscripts described by the Brazilian regulatory agency (Ministry of Education of Brazil) indicate that the word ethics appears 9 times. Moreover, the Brazilian government documents are clear about software engineers' skills required to make sound ethical choices: ``To be able to investigate, understand and structure the features of application domains in diverse contexts considering ethical, social, legal and economic issues, individually and/or as a team." 

%In the Table \ref{table:ethics_courses} we summarize the percent of mandatory Ethics -related course workload comparing with the total program work load for the top 5 Brazilian SE Degree Programs.
%\begin{table}[ht]
%\renewcommand{\arraystretch}{1.3}
%\caption{Percent of Ethics-related Course in the SE Program}
%\label{table:ethics_courses}
%\centering
%\begin{tabular}{|c|c|c|c|c|}
%\hline
%UNB & UFG & UEPG & UFC & UNIPAMPA\\
%\hline
%1,72\% & 2,13\% & 1,44\% & 2,08\% & 0,00\%\\
%\hline
%\end{tabular}
%\end{table}

Since undergraduate degree programs in Software Engineering are relatively new in Brazil and there are few quality indicators for a more accurate selection, we focused our research on the top two Brazilian Software Engineering programs (5 Stars) according the Student Guide 2016 \cite{guia}, the most popular university ranking guide of the nation. So, the study sample consists of analyzing the pedagogical project of these courses: University of Bras\'{\i}lia (UNB) and Federal University of Goi\'{a}s (UFG). The analysis was mainly based on searching for the word ethics on the course syllabi.  

The SE program of the University of Bras\'{\i}lia has a work load of 3,480 hours and its undergraduate students must be able to ``be oriented to act as a social transformer, aiming at social welfare and ethically evaluating the social and environmental impact of their interventions." However, there is only one compulsory ethics-related course named ``Humanity and Citizenship" with 60 hours, which corresponds to 1.72\% of the total content. This course aims to ``Introduce the concepts of humanities, social sciences and citizenship to foster the critical view and awareness of the humanistic, social, ethnic-racial, political, economic, ethical and environmental issues involved in the professional action of the engineer." There are also two other optional courses that reflect an ethics perspective: ``Information, Communication and the Knowledge Society" and ``Productivity and Professionalism in Software Engineering". 

The SE program of the Federal University of Goi\'{a}s has a work load of approximately 3,000 hours and, according the pedagogical project, the former students must be able to ``develop an active and ethical posture". However, there is only one mandatory Ethics-related course named ``Ethics, Norms, and Professional Posture" being 64 hours in length, which corresponds only to 2.13\% of the total content. This course introduces the following topics: ``Notions of ethics. A Code of ethics for software engineers. Overview of international norms and standards, laws and local resolutions relevant to Software Engineering. Nomenclature used by the area according to IEEE Std. 12207-2008. Conflict resolution. How to prepare for and behave in meetings. Hygiene aspects. Presentation aspects. Aspects of conduct. Entrepreneurial attitudes. Entrepreneur instruments (business plan and others). Techniques for identifying opportunities and procedures for opening a business", which contains many other unrelated topics.

\begin{table*}[h]\centering
\ra{0.7}
\caption{Mapping technical components to SE Curricula}
\label{tab:techmap}
\footnotesize
\begin{tabular}{@{}L{3cm} L{5.5cm} L{8.5cm}@{}}
\toprule
{\textbf{Cybertechnologies (technical components)}} & {\textbf{Related ACM/IEEE SEEK units (hours)}} & {\textbf{Related disciplines of top 2 BR SE Programs (hours)}}                     \\ \midrule
Data Science & 
Modeling foundations (8); Design strategies (6); Detailed design (14); 
&  UFG - Database Systems (64); Software Development for Persistence (64); Detailed Software Design (64).  \linebreak     UNB - Database Systems I (60); Database Systems II (60). 
\\ \midrule
Cloud Computing & Construction technologies (20);                Architectural design (12);
Detailed design (14); & UFG - Concurrent Software Development (64); Networks and Distributed Systems (64); Operating Systems (64). \linebreak UNB - Programming for Parallel and Distributed Systems (60); Computer Network Fundamentals (60); Operating Systems Fundamentals (60).                           
\\ \midrule
Algorithms & Computer science foundations (120); 
Problem analysis and reporting (5); Analysis fundamentals (8). 
& UFG - Introduction to Programming (64); Algorithms Fundamentals and Data Structures (64); Software Development for Web (64). \linebreak 
UNB - Data Structures II (60); Analysis of Algorithms (60); Techniques of Programming in Emerging Platforms (60).             \\ \midrule
Artificial Intelligence; Robotics                                 & Construction technologies (20); Construction tools (12); 
Architectural design (12); Engineering foundations for software (22). 
& UFG - Computer Architecture (64); Software Architecture (64).   \linebreak 
UNB - Embedded Systems Fundamentals (60); Theory of Digital Electronics I (60); Computer Architecture Fundamentals (60).     \\ \midrule
Internet; Internet of Things; Mobile; Social Media               & Construction technologies (20); Architectural design (12); 
Detailed design (14);
Computer and network security (8);
Introduction to Computer Systems (60).  
& UFG - Software Architecture (64); Integration of Applications (64); Software Development for Devices (64). \linebreak 
UNB - Software Architecture (60); Embedded Systems Fundamentals (60). 
\\ \midrule
Secure Software Engineering  & Security fundamentals (4); Computer and network security (8); Developing secure software (8).
& UFG - Secure (64); Networks and Distributed Systems (64).       
\linebreak UNB - Computer Network Fundamentals (60); Operating Systems Fundamentals (60).                                       
\\ \bottomrule                                    
\end{tabular}
\end{table*}

\section{Integrating cyberethics into the SE curriculum}
% adapted from A Multi-level Perspective for the Integration of Ethics, Corporate Social Responsibility and Sustainability (ECSRS) in Management Education
\subsection{Institutional level}
Software engineering professors are in the best position to spark a dialogue about being a professional in a community that shares the same weighty responsibilities \cite{Narayanan2014}. They, therefore, must be prepared to conduct a controversial and relevant dialogue with students. Professors need to collaborate with ethicists, complex systems scholars, professional societies and a number of other discipline representatives in order to keep themselves updated.

Therefore, the institution must create the necessary conditions to foster this community. Cooperation among software engineers, corporations, social scientists, ethicists, philosophers, faith leaders, economists, lawyers, and policymakers will shape cyberethics' policies for the 21st century. 

Institutions need to establish an ethical-oriented culture, incorporating cyberethics in their strategic plans, allocating resources for related initiatives, and designing curricular and extracurricular activities for all students, faculty and staff \cite{Pamies2016}.

\begin{table*}[t!h]\centering
\ra{0.7}
\caption{Mapping non-technical components to SE Curricula}
\label{tab:nontechmap}
\footnotesize
\begin{tabular}{@{}L{3cm} L{6.3cm} L{7.7cm}@{}}
\toprule
{\textbf{Cybertechnologies (non-technical components)}} & {\textbf{Related ACM/IEEE SEEK units (hours)}} & {\textbf{Related disciplines of top 2 BR SE Programs (hours)}}                     
\\ \midrule
Experience, Product and Services Design; Design Thinking & Eliciting requirements (10); Types of models (12); Process concepts (3); Design strategies (6).
& UFG - Software Requirements (64); Human-computer Interaction (64); Software Engineering Methods (64). \linebreak UNB - Software Requirements (60); Human-computer Interaction (60); Software Development Methods (60).                       \\ \midrule
Software and Society; Cyberlaw                                    & Software quality concepts and culture (2); Group dynamics and psychology (8);  Security fundamentals (4); Professionalism (6).    
& UFG - Introduction to Software Engineering (96); Ethics, Norms and Professional Posture (64).                                   \linebreak UNB - Software Product Engineering (60); Engineering and Environment (60); Humanity and Citizenship (60).             \\ \midrule
Digital Business Models; Economics                               & Engineering economics for software (8); Requirements fundamentals (6); 
Types of models (12);
& UFG - Software Requirements (64); Software Economic Engineering (64). \linebreak UNB - Software Requirements (60); Economic Engineering (60).                        \\  \bottomrule                                   
\end{tabular}
\end{table*}

% create online case studies that emphasize ethical decision making in both
% individual/professional and broader societal contexts (including cases that
% examine the implications for engineering practice of corporate social
% responsibility programs);
% • Collaboration among engineering and computing ethicists to identify, expand
% or modify where appropriate, and publicize existing online computer ethics
% resources with significant relevance to engineering education and practice; and
% • Collaboration among ethicists, engineering educators, and professional
% engineering societies to create explicit, transparent links between public policy
% positions advocated by professional societies and relevant principles espoused
% in their codes of ethics.
% Ways of Thinking about and Teaching Ethical Problem Solving: Microethics and Macroethics in Engineering. Joseph R. Herkert 2005

\subsection{Curricular level}
At the curricular level, we need to consider that different strategies should be considered in designing the academic curricula to integrate cyberethics content. The incorporation of content in cyberethics through \textit{stand-alone subjects} (specific courses) or \textit{embedded subjects} (different courses) constitutes one of the principal current debates of how to teach ethics\cite{Pamies2016}. 

%SE educators include a discussion of ethics with every significant technology they teach. 
Based on related studies \cite{Holland2013, Narayanan2014}, we recommend the embedded subjects approach, as knowledge from different disciplines can allow for a more holistic understanding of controversial cyberethics issues. This means that cyberethics should be integrated into different SE courses. Stand-alone disciplines can be added to guarantee the commitment of a certain number of credit hours in the curriculum.

Following the proposed conceptual framework, we suggest a mapping of cybertechnology technical (Table \ref{tab:techmap}) and non-technical components (Table \ref{tab:nontechmap}) to the ACM/IEEE SEEK units and to the disciplines of the top two Brazilian SE Programs, also expressing their workload. It is possible to observe that these components have an interdisciplinary approach, increasing the number of courses in which cyberethics debates can take place. 

Adding cyberethics content in the ACM/IEEE SEEK expands the ethics program to 66.8\% of the total content, instead of the previous 1.3\% (Professionalism- 6 hours), and in the top two Brazilian SE Programs extend the ethics-related topics on average to 37.8\%, instead of previous 1.9\% (Ethics, Norms and Professional Posture - 64 hours and Humanity and Citizenship - 60 hours). Regarding content, we recommend that professors pursuing integrating cyberethics into their disciplines analyze the most prominent controversies in cybertechnologies from our conceptual framework (Table \ref{tab:tech}). 

\subsection{Instrumental level}
At this level, instructors need to set specific learning objectives and decide on the pedagogical tools to be used. A number of methodologies can be used. In this phase, we confirm research findings and recommendations on assessing millennial generation's learning preferences and learning styles \cite{BenJacob2005, Wilson2008, Farrell2014, Narayanan2014}. We thus provide some concrete advice based on these studies and our own teaching experience.

According to Wilson and Gerber \cite{Wilson2008}, millennials are decidedly active learners. They have a hypertext mindset, which diminishes the applicability of a lecture-style training format. They are also multitaskers with a propensity for innovation fueled by curiosity, discovery, and exploration. Factors such as shorter attention spans, low boredom tolerance, and necessity of hands-on elements contribute to the millennial generation's active learning style.

Millennials expect to have choices, so the learning process should provide opportunities for creating their own learning or meaning within courses, a form of active involvement through self-tailoring. Case studies, hypotheticals, role-playing, storytelling, simulations, journaling, activity logs, and teaching approaches are common tools of contemporary education that lend to millennials' preference for tailored classes. 

Finally, common feature of contemporary classes involves decreasing the amount of content from courses, because students have greater and continuous access to content. Content-mastery is less crucial than thoughtful processing and critical analysis. This could be achieved by a deeper exploration of materials.

\section{Conclusion}
%Embedding coverage of ethics in software engineering courses would help students draw strength and wisdom from dialogue with other future members of their profession colleagues who will face the same types of moral dilemmas, struggle with the same sorts of tough decisions, and ultimately seek to earn, in similar ways, the respect of their peers and the broader public.\cite{Narayanan2014}

Millennial Software Engineers must think and act ethically in order to deal with critical issues of the new century. However, there is still open challenges in the educational systems and curricula delivery since ethics concerns are presented in isolated courses. This study presented a conceptual framework for cyberethics and conducts an analysis of the SE curriculum proposed by the ACM/IEEE SE Curriculum Guidelines, as well as the top Brazilian SE undergraduate programs. Suggestions on how to introduce this theme as an interdisciplinary curriculum approach are also provided.

This work points to the need for some interesting future studies. The next step is to more thoroughly investigate cyberethics education concrete outcomes by administering a survey to professors and students. Another important area involves evaluating additional SE curriculum guidelines from different countries and investigating cultural factors influencing cyberethics. Finally, future studies may aim to determine a strategy to introduce cyberethics in any SE-related course as a fundamental feature of millennial software engineers instruction.

\Urlmuskip=0mu plus 1mu\relax
\bibliographystyle{IEEEtran}
{\footnotesize
\bibliography{IEEEabrv,icse}}


%\end{thebibliography}

\end{document}